\documentclass[useAMS,usenatbib,usegraphicx]{mn2e}

\newcommand{\be}{\begin{equation}}
\newcommand{\ba}{\begin{eqnarray}}
\newcommand{\ee}{\end{equation}}
\newcommand{\ea}{\end{eqnarray}}
\newcommand{\etal}{et al.\ }
\newcommand{\etalb}{et al.}

\newcommand{\Omm}{\Omega_m}

\newcommand{\Lya}{\mbox{Ly$\alpha$} }

\newcommand{\del}{\delta}

\title[A Model For Infall Around Virialized Halos]{A Model 
For Infall Around Virialized Halos}
\author[R. Barkana]{R. Barkana$^{1}$\thanks{E-mail:
barkana@wise.tau.ac.il}\\
$^{1}$School of Physics and Astronomy, Tel Aviv University, Tel Aviv 69978,
ISRAEL}
\begin{document}

\pagerange{\pageref{firstpage}--\pageref{lastpage}} \pubyear{2002}

\maketitle

\label{firstpage}

\begin{abstract}
Motivated by the recent direct detection of cosmological gas infall,
we develop an analytical model for calculating the mean density
profile around an initial overdensity that later forms a dark matter
halo. We account for the problem of peaks within peaks; when
considering a halo of a given mass we ensure that this halo is not a
part of a larger virialized halo. For halos that represent high-sigma
fluctuations we recover the usual result that such halos
preferentially lie within highly overdense regions; in the limit of
very low-sigma fluctuations, on the other hand, we show that halos
tend to lie within voids. Combined with spherical collapse, our
results yield the typical density and velocity profiles of the gas and
dark matter that surround virialized halos.
\end{abstract}


\section{Introduction}

\label{sec:Intro}

Numerical simulations of hierarchical halo formation indicate a
roughly universal spherically-averaged density profile within
virialized dark matter halos \citep{NFW}. Such profiles are generated
by complicated processes that involve violent relaxation and multiple
shell crossing; the great interest in them results from the
possibility of a direct comparison to the profiles inferred from
galactic rotation curves \citep[e.g.,][]{sb00,dmr01,vs01} and from
lensing in clusters \citep[e.g.,][]{ste02,dhs02}. However, the
analogous question of what is the typical density profile outside the
virial radius has received relatively little attention, because of the
difficulty of directly observing density profiles even out to the halo
virial radius; this difficulty results from the rapid decrease in the
gas and dark matter densities with distance from the halo
center. However, the theoretical description is simpler for infall
outside the virialization radius, since in cold dark matter (hereafter
CDM) models there is no shell crossing until the final non-linear
collapse.

The gas around halos can only be detected using physical effects that
are very sensitive to low-density gas. One such phenomenon is resonant
\Lya absorption; \citet{GP} first noted that even an H I neutral fraction 
$\sim 10^{-5}$ suffices to make gas at the cosmic mean density
strongly absorb photons at the \Lya resonance. Thus, the pattern of
gas infall around halos may affect observable properties of the \Lya
absorption in these regions. At moderate redshifts, these regions near
halos play a prominent role in measurements of the proximity effect of
quasars \citep[e.g.,][]{prox1, prox2}. In these measurements, the
effect of the high quasar intensity on the statistics of \Lya
absorption is analyzed and used to infer the mean ionizing background,
since the quasar intensity dominates only as long as it is
significantly stronger than the background. A different application
may be possible at very high redshifts, corresponding to the time
before the universe was fully reionized; the presence of the fully
neutral IGM could be detected through absorption of a bright source by
the \Lya damping wing of the IGM \citep{jordi98}. However, every
bright source creates a surrounding H II region which complicated the
interpretation \citep{CH00,MR00}. The properties of the gas in the
regions that lie just outside the host halo of the source galaxy or
quasar should play a major role in these various applications of the
observed absorption.

The various calculations of absorption cited above neglected the
enhanced densities and the negative (i.e., infall) velocities expected
for the gas surrounding a massive halo at high redshift. This infall,
however, may be an essential component of any correct interpretation
of observations; \citet{le95} applied it to the proximity effect and
showed that neglecting infall can lead to an overestimate of the
ionizing background flux by up to a factor of three. Applying the same
infall profile, \citet{nature} have recently revealed the signature of
infalling gas around high redshift quasars. The spectral pattern due
to \Lya absorption by the infalling gas can be used to estimate the
quasar's ionizing intensity and the total mass of its host halo. While
the spectra of individual objects are marked by the large density
fluctuations expected in the intergalactic medium (hereafter IGM), if
this pattern is averaged over many objects then the resulting mean
profile could be compared directly with an accurate theoretical
prediction for the density and velocity profile of the infalling gas.

\citet{le95} defined a halo in the same way as halos were defined in the 
classic model of \citet{PS}. In this approach, halos on a given scale
are defined as forming when the initial density averaged on this scale
is higher than a threshold that depends on the collapse redshift and
is fixed using spherical collapse. The halo mass function derived from
this assumption, when multiplied by an ad-hoc correction factor of
two, provides a reasonable match to the results of numerical
simulations \citep[e.g.,][]{katz93}.

The same mass function was later derived by \citet{bc91} using a
self-consistent approach, with no need for external factors. In their
derivation, the factor of two has a more satisfactory origin, namely
the so-called ``cloud-in-cloud'' problem: if the average initial
density on a certain scale is above the collapse threshold, this
region may be contained within a larger region that is itself also
denser than the collapse threshold; in this case the original region
should be counted as belonging to the halo with mass corresponding to
the larger collapsed region. \citet{bc91} and \citet{PS} both give the
same halo mass function and thus appear to be at least statistically
valid given the agreement with the mass function seen in numerical
simulations. However, the \citet{bc91} model is more satisfactory
since it accounts for the cloud-in-cloud problem and, more
importantly, allows for calculations of other halo properties in
addition to simple mass functions, all in an approach that
is more self-consistent.

The \citet{bc91} approach (also known as extended Press-Schechter) has
been used to calculate halo merger histories \citep{lc93}, halo mass
functions in models with warm dark matter \citep{wdm}, halo mass
functions in a model with ellipsoidal collapse \citep{ellipse}, and
the nonlinear biasing of halos \citep{porciani,evan}. In this paper,
we present a new application of this approach in which we obtain the
expected profile of infalling matter around virialized halos. The rest
of this paper is organized as follows. In \S~2.1 we establish our
notation and review the \citet{bc91} derivation of the \citet{PS} halo
mass function. In \S~2.2 we review the infall profile used by
\citet{le95} and also derive an alternative model based on similar
assumptions but carried out in Fourier space. In \S~2.3 we derive our
model based on the extended Press-Schechter formalism. In \S~2.4 we
then discuss the modified picture of halo virialization once infall is
included. In \S~3 we illustrate the predictions of our model for the
initial and final density profiles surrounding halos. Finally in
\S~4 we give our conclusions.

\section{A Model for Infall: Derivation}

\subsection{Review of Halo Collapse}

\label{sec:collapse}

Before addressing the problem of the initial density profile around a
virialized halo, we briefly review in this section the approach of
\citet{bc91} which leads to the halo mass function. We work with the
linear overdensity field $\del({\bf x},z) \equiv
\rho({\bf x},z)/\bar\rho(z) - 1$, where ${\bf x}$ is a comoving
position in space, $z$ is the cosmological redshift and $\rho$ is the
mass density, with $\bar \rho$ being the cosmic mean density. In the
linear regime, the density field maintains its shape in comoving
coordinates and the overdensity simply grows as $\del =
\del_i D(z)/D(z_i)$, where $z_i$ and $\del_i$ are the initial redshift
and overdensity, and $D(z)$ is the linear growth factor
\citep{p80}. When the overdensity in a given region becomes
non-linear, the expansion halts and the region turns around and
collapses to form a virialized halo.

The time at which the region virializes can be estimated based on the
initial linear overdensity, using as a guide the collapse of a
spherical tophat perturbation. At the moment at which a tophat
collapses to a point, the overdensity predicted by linear theory is
$\del_c=1.686$ \citep{p80} in the Einstein-de Sitter model. This value
depends slightly on cosmological parameters, and even in the
Einstein-de Sitter case it is modified by infall, as we discuss in \S
\ref{sec:deltav}; our analytical expressions in this section are
valid for any value of $\del_c$.

A useful alternative way to view the evolution of density is to
consider the linear density field extrapolated to the present time,
i.e., the initial density field at high redshift extrapolated to the
present by multiplication by the relative growth factor. In this case,
the critical threshold for collapse at redshift $z$ becomes redshift
dependent, \be \del_c(z) = \del_{c} / D(z)\ .\ee We adopt this view,
and throughout this paper the power spectrum $P(k)$ refers to the
initial power spectrum, linearly extrapolated to the present (i.e.,
not including non-linear evolution).

At a given $z$, we consider the smoothed density in a region around a
fixed point $A$ in space. We begin by averaging over a large mass
scale $M$, and then lower $M$ until we find the highest value for
which the averaged overdensity is higher than $\del_c(z)$; we assume
that the point $A$ belongs to a halo with a mass $M$ corresponding to
this filter scale.

In this picture we can derive the mass distribution of halos at a
redshift $z$ by considering the statistics of the smoothed linear
density field.  If the initial density field is a Gaussian random
field and the smoothing is done using sharp $k$-space filters, then
the value of the smoothed $\del$ undergoes a random walk as the cutoff
value of $k$ is increased. If the random walk first hits the collapse
threshold $\del_c(z)$ at $k$, then at a redshift $z$ the point $A$ is
assumed to belong to a halo with a mass corresponding to this value of
$k$. Instead of using $k$ or the halo mass, we adopt as the
independent variable the variance at a particular filter scale $k$,
\be
S_k \equiv \frac{1}{2 \pi^2} \int_0^k dk'\, k'^2\, P(k')\ .
\label{eq:Sk}
\ee

In order to construct the number density of halos in this approach, we
need to solve for the evolution of the probability distribution
$Q(\del,S_k)$, where $Q(\del,S_k)\, d\del$ is the probability for a
given random walk to be in the interval $\del$ to $\del+d\del$ at
$S_k$. Alternatively, $Q(\del,S_k)\, d\del$ can also be viewed as the
trajectory density, i.e., the fraction of the trajectories that are in
the interval $\del$ to $\del+d\del$ at $S_k$, assuming that we
consider a large ensemble of random walks all of which begin with
$\del=0$ at $S_k=0$.

\citet{bc91} showed that $Q$ satisfies a diffusion equation,
\be
\frac{\partial Q}
{\partial S_k} = \frac{1} {2} \frac{\partial^2 Q} {\partial \del^2},
\label{eq:oneddiff}
\ee
which is solved by the Gaussian solution which we label $Q_0$:
\be
Q_0(\del,S_k) = \frac{1}{\sqrt{2 \pi S_k}} \exp \left[
-\frac{\del^2}{2\, S_k} \right]\ .
\ee

To determine the probability of halo collapse at a redshift $z$, we
consider the same situation but with an absorbing barrier at
$\del=\nu$, where we set $\nu=\del_c(z)$. The fraction of trajectories
absorbed by the barrier up to $S_k$ corresponds to the total mass
fraction contained in halos with masses higher than the value
associated with $S_k$. In this case, $Q$ again satisfies
eq.~(\ref{eq:oneddiff}) and the solution with the barrier in place is
given by adding an extra image-solution:
\be
Q(\nu,\del,S_k)= Q_0(\del,S_k)-Q_0(2 \nu-\del,S_k) .
\label{eq:1}
\ee
Using this expression, the fraction of all trajectories that have
passed above the barrier $\nu$ by $S_k$ is
\be
F(\nu,S_k) = 1-
\int_{-\infty}^{\nu}  d\del\, Q(\nu,\del,S_k)\ ,
\ee
and the differential mass function is found [using
eq.~(\ref{eq:oneddiff})] to be
\be 
f(\nu,S_k) =
\frac{\partial }{\partial S_k} F(\nu,S_k) = 
\frac{\nu }{\sqrt{2 \pi} S_k^{3/2}} \exp\left[-\frac{\nu^2}{2 S_k}
\right]\ . \label{eq:f1pt} \ee 
As $f(\nu,S_k)\, dS_k$ is the probability that point $A$ is in a halo
with mass in the range corresponding to $S_k$ to $S_k+d S_k$, the halo
abundance is then simply
\be
\frac{dn}{dM} = \frac{\bar{\rho}}{M} \left|\frac{d S_k}{d M} \right|
f(\nu,S_k)\ ,
\label{eq:abundance}
\ee
where $dn$ is the comoving number density of halos with masses in the
range $M$ to $M+dM$. The cumulative mass fraction in halos above mass
$M$ is similarly determined to be
\be
\label{eq:PSerfc} F(>M | z)={\rm erfc}\left(\frac{\nu}
{\sqrt{2 S_k}} \right)\ .
\label{eq:fm1point}
\ee

While these expressions were derived in reference to density
perturbations smoothed by a sharp $k$-space filter as given in
eq.~(\ref{eq:Sk}), $S_k$ is often replaced in the final results with
the variance of the mass $M$ enclosed in a spatial sphere of comoving
radius $r$:
\be
\sigma^2(M) = \sigma^2(r) = \frac{1}{2 \pi^2} \int_0^\infty k^2 dk P(k) 
W^2(k r)\ ,
\label{eq:sig}
\ee
where $W(x)$ is the spherical tophat window function, defined in
Fourier space as
\be
W(x) \equiv 3 \left[ \frac{\sin(x)}{x^3} - \frac{\cos(x)}{x^2} \right]\ ,
\label{eq:reW}
\ee 
and $M=4 \pi \bar \rho(0) r^3/3$ in terms of the mean density of
matter at $z=0$. With these replacements we recover the cumulative
mass fraction that was originally derived \citep{PS} simply by
considering the distribution of overdensities at a single point,
smoothing with a tophat window function, and integrating from
$\delta_c$ to $\infty$.  In this derivation the authors were forced to
multiply their result by an arbitrary factor of two, to account for
the mass in underdense regions. The excursion-set derivation presented
here, based on \citet{bc91}, properly includes all the mass by
accounting for small regions that lie within overdensities on larger
scales. This approach also makes explicit the approximations involved
in working with $\sigma^2(r).$ Strictly speaking, dealing with a
real-space filter requires a complete recalculation of
$f(\nu,\sigma^2)$ which accounts for the correlations intrinsic to
$W(x)$. However, simply replacing $S_k$ with $\sigma^2(r)$ in
eq.~(\ref{eq:abundance}) has been shown to be in reasonable agreement
with numerical simulations
\citep[e.g.,][]{katz93}, and is thus a standard approximation.

\subsection{Infall based on the Press-Schechter Model}

In the Press-Schechter model, a halo of mass $M$ is assumed to form at
redshift $z$ if the corresponding comoving sphere (of radius denoted
$r_M$) has a linear overdensity of $\delta = \nu \equiv
\delta_c(z)$. \citet{le95} used this fact to derive a mean infall
profile as follows. Consider spheres of various comoving radii $r$,
all centered on the region in the initial density field that contains
the mass that ends up in the final virialized halo. The mean enclosed
overdensity $\delta(r)$ in each such sphere (linearly extrapolated to
the present) is a Gaussian variable with variance $\sigma^2(r)$ as
given in eq.~(\ref{eq:sig}). The overdensities in two such spheres
with radii $r_1$ and $r_2$ are correlated with a cross-correlation
\be
\xi_r(r_1,r_2) \equiv \frac{1}{2 \pi^2} \int_0^\infty k^2 dk
P(k) W(k r_1) W(k r_2)\ .
\label{eq:xir}
\ee
Thus, the expected value of $\delta(r)$ given that $\delta_M \equiv
\delta(r_M)=\nu$ is given by their joint Gaussian distribution as
\be \frac{\langle\delta(r)\rangle_{\rm PS-r}} {\nu} = \frac{\xi_r(r_M,r)}
{\sigma^2(r_M)}\ , \label{eq:PSr} \ee where the notation $PS-r$
indicates a derivation based on the Press-Schechter model and carried
out in $r$-space.

In addition to obtaining the mean profile, it is useful to calculate
the scatter in the initial profiles around halos of a given mass in
order to assess the applicability of the mean profile to individual
objects. If in the future the model is to be compared to halos in
numerical simulations or in real observations, a quantitative formula
for the scatter helps us determine the number of objects necessary to
average over before the mean profile emerges. In addition, the scatter
itself is a prediction of the model that in principle can be compared
to simulations or observations. In the present model, the ratio
$\delta(r) / \nu$ follows a Gaussian distribution with the mean value
given above and a standard deviation given by
\be \Sigma(r)_{\rm PS-r} = \sqrt{\frac{\sigma^2(r)}{\nu^2} - 
\frac{\xi_r^2(r_M,r)}{\nu^2 \sigma^2(r_M)}} \ . \ee

Alternatively, we can create an infall model based on similar
assumptions, but (as in the previous subsection) calculated in
$k$-space. Consider a halo that forms on a scale with a corresponding
variance $S_{k,M}$, and consider a larger scale around it with
variance $S_k < S_{k,M}$. We obtain here a Press-Schechter model in
$k$-space by considering random trajectories but without including an
absorbing barrier; the self-consistent inclusion of a barrier, which
accounts for peaks-within-peaks and is the hallmark of the extended
Press-Schechter model, is considered in the next subsection.

In the absence of a barrier, the probability distribution of the
overdensity $\delta$ at any $S_k$ is Gaussian, and the values of
$\delta$ at multiple points are jointly normal. Now, $\delta$ has
variance $S_k$ and $S_{k,M}$ at the two scales of interest.
Furthermore, since the random walk from $S_k$ to $S_{k,M}$ is
independent of the random walk from 0 to $S_k$, the cross-correlation
of the values of $\delta$ at $S_k$ and at $S_{k,M}$ is given by the
overlapping portion of the two random walks: the cross-correlation is
therefore simply $S_k$. Similarly to the $r$-space case above, the
joint Gaussian distribution yields
\be \frac{\langle\delta(r)\rangle_{\rm PS-k}} {\nu} = \frac{S_k} 
{S_{k,M}}\ . \label{eq:PSk} \ee As in the PS-r model, the distribution
of $\delta(r) / \nu$ is Gaussian, where now the standard deviation is
\be \Sigma(r)_{\rm PS-k} = \sqrt{\frac{S_k}{\nu^2} - 
\frac{S_k^2}{\nu^2 S_{k,M}}} \ . \ee

Note that both of the models developed in this section satisfy the
continuity condition $\langle\delta(r)\rangle/\nu \rightarrow 1$ when
$r \rightarrow r_M$ (or equivalently $S_k \rightarrow
S_{k,M}$). However, we show in the next section that they fail to give
a physically acceptable result in a different limit, when $\nu \ll
\sqrt{S_{k,M}}$.  Moreover, we have ignored in this section the
failure of the simple Press-Schechter approach to account for negative
fluctuations, and have not included the factor-of-two normalization
fudge factor; in the next section, in contrast, we develop a model
that is self-consistent.

\subsection{Infall based on the Extended Press-Schechter Model}

In this section we derive a new model for the initial density profile
around a virialized halo. As additional motivation, we first show that
the two simple models considered in the previous section yield results
that are physically unacceptable, once we consider the obvious
possibility that a small halo will lie within a bigger one. Consider,
for example, the PS-r model in a CDM cosmology (the cosmological
parameters are listed in section \ref{sec:res}). In order to explore
the limit of $\nu \ll \sigma(r_M)$ at $z=0$ (when $\nu=1.68$ in this
model), we consider a halo of mass $M=10^6 M_{\odot}$ ($r_M=18$ kpc,
$\sigma(r_M)=8.5$), and a surrounding region with $r = 2 r_M$ ($M(r)=8
\times 10^6 M_{\odot}$, $\sigma(r)=7.4$). Consider randomly chosen
regions of size $r$ in the universe. Half of them will have a negative
overdensity and half will have a positive overdensity (where, as
elsewhere in this paper, we consider the linearly-extrapolated
overdensity). Those with a positive perturbation will almost certainly
be contained within a halo of mass at least $M(r)$, since the typical
$\delta$ on this scale is of order $\sigma(r)$, which in turn is much
greater than the threshold $\nu$ for forming a halo. Even for those
rare perturbations that are smaller than $\nu$, the fact that the
perturbation is positive increases the (already very high) chance for
the formation of some halo with radius between $r_M$ and $r$. Clearly,
among regions with positive perturbations on the scale $r$, very few
will contain an isolated halo of size $r_M$ that is not contained
within a larger halo.

On the other hand, fully half of such large regions have a negative
perturbation, again with magnitude of order $\sigma(r)$. Such a
negative perturbation implies that certainly no halo has formed on the
scale $r$; together with the correlation among spheres of different
radii, it also implies a very strong reduction in the chance that any
halo has formed on a scale between $r_M$ and $r$. This double effect
implies that isolated halos of size $r_M$ will be found almost
exclusively within voids on the scale $r$. Indeed, in the limit of
$\nu / \sigma(r_M) \rightarrow 0$, $\delta$ on the scale $r$ should be
negative with a magnitude at least comparable to $\sigma(r_M)$, in
order for $\delta$ to remain negative at all scales from $r$ down to
$r_M$. Thus, in this limit we should find that the ratio
$\delta(r)/\nu \rightarrow -\infty$. This contrasts with the
Press-Schechter models from the previous section, which predict that
the typical such case should have a positive $\delta$ on the scale $r$
(in fact, a $\delta$ only slightly below $\nu$); this behavior arises
from the general failure of these models to properly incorporate
negative density fluctuations.

To derive our new, self-consistent model, we apply the extended
Press-Schechter model in Fourier space and then use an ansatz to
convert it to real space.  We begin by following the derivation of
$\langle\delta(r)\rangle_{\rm PS-k}$ in the previous subsection but
now including the absorption barrier at $\delta=\nu$. Thus, we
consider once again the two scales with variances $S_{k,M}$ and $S_k$,
but we now require $S_{k,M}$ to form a halo that is not contained
within any larger virialized halo; in particular this implies that
there is no halo on the scale corresponding to $S_k$ or on any larger
scale. Then the probability distribution of the overdensity $\delta$
at $S_k$ is given by $Q(\nu,\del,S_k)$ as in eq.~(\ref{eq:1}). As
before, the random walk from $S_k$ onwards is independent of the
random walk leading up to $S_k$. In order to account for the extended
Press-Schechter definition of halo formation, we use arguments similar
to those leading to eq.~(\ref{eq:f1pt}) [see \citet{bc91} and
\citet{lc93} for other applications of similar arguments]; the
probability distribution of the initial density at $S_k$, given the
formation of a halo at $S_{k,M}$, is found to be \be P(\delta |
\delta_M=\nu)_{\rm ePS-k} = Q(\nu,\del,S_k)\, \frac{f(\nu-\del, 
S_{k,M}-S_k)} {f(\nu,S_{k,M})}\ . \label{eq:pPSk} \ee The mean
expected $\delta$ is given by integrating $\delta$ over this
distribution, with limits of $-\infty$ to $+\nu$; we thus obtain the
result of using the extended Press-Schechter model in $k$-space:
\ba \frac{\langle\delta(r)\rangle_{\rm ePS-k}} {\nu} & = & 1 - \left(1- 
\alpha +\frac{\alpha}{\beta} \right)\, {\rm erf}\left[ \sqrt{\frac
{\beta(1-\alpha)}{2 \alpha}} \right] \nonumber \\
& & - \sqrt{ \frac{2\alpha(1 - \alpha)} {\pi \beta}}\, \exp \left[ - 
\frac {\beta(1-\alpha)}{2 \alpha} \right] \ , \label{eq:ePSk} \ea
where we have defined
\be \alpha = \frac{S_k} {S_{k,M}}\ \ ;\ \ \beta = \frac{\nu^2}
{S_{k,M}}\ . \label{eq:alpha} \ee

Unlike the models in the previous subsection, the present model does
not yield a Gaussian probability distribution for the density. 
To find the scatter, we first integrate the probability
distribution given by eq.~(\ref{eq:pPSk}) and find the cumulative
probability distribution $P_C$. Using the variable $\gamma=\delta/\nu$
instead of $\delta$, the probability of obtaining a value of
$\delta(r)$ that is between $-\infty$ and $\nu \gamma$ equals 
\ba P_C(\gamma)_{\rm ePS-k} & = & 1 + \frac{1}{2} \left[
{\rm erf}(A)+{\rm erf}(B) \right] \nonumber \\ & & +
\sqrt{\frac{\alpha}{2 \pi \beta (1-\alpha)}}\, \left[ e^{-A^2}-
e^{-B^2} \right]\ , \label{eq:Pcum} \ea
where we use the shorthand
\be A = \frac{(\gamma-\alpha)\sqrt{\beta}} {\sqrt{2 \alpha (1-\alpha)}}\ \ ;\ 
\ B = \frac{(\gamma+\alpha-2)\sqrt{\beta}} {\sqrt{2 \alpha (1-\alpha)}}\ . 
\label{eq:Pcum2} \ee
We can now numerically calculate the $1-\sigma$ interval of
$\delta(r)$ by bracketing the central $68.3\%$ of the probability.
For a given $M$ and $\nu$, we thus define the minus $1-\sigma$ density
profile as given by values of $\gamma$ determined at each $r$ by
$P_C(\gamma)= 15.9 \%$, while the plus $1-\sigma$ profile is
determined by $P_C(\gamma)= 84.1 \%$.

Having developed several possible infall models, we compare their
relative merits by considering several limiting cases.  First, when $r
\rightarrow r_M$ (or equivalently $S_k \rightarrow S_{k,M}$), all the
models satisfy the continuity condition $\langle\delta(r)\rangle/\nu
\rightarrow 1$ with zero scatter. In the limit of $\beta \rightarrow 0$, 
we require $\langle\delta(r)\rangle/\nu \rightarrow -\infty$ as
explained earlier in this section. While $\langle\delta(r)\rangle_{\rm
ePS-k}$ behaves correctly in this limit (and approaches a constant
that does not depend on $\nu$), $\langle\delta(r)\rangle_{\rm PS-r}$
and $\langle\delta(r)\rangle_{\rm PS-k}$ are proportional to $\nu$ and
thus fail to correctly describe the $\beta \rightarrow 0$ limit. Thus,
only $\langle\delta(r)\rangle_{\rm ePS-k}$ behaves physically in both
of the limits $r \rightarrow r_M$ and $\beta \rightarrow 0$.

In the limit of a halo that represents an extremely rare peak, i.e.,
$\beta \rightarrow \infty$, the Press-Schechter model becomes an
accurate description that is equivalent to the extended
Press-Schechter model; the reason is that if the central halo is
extremely rare then it is exceedingly improbable to find a virialized
halo on any scale larger than $r_M$, and thus it is acceptable to
neglect the barrier at $\delta=\nu$ except right near the scale
$r_M$. In this limit, $\langle\delta(r)\rangle_{\rm ePS-k}$ and
$\langle\delta(r)\rangle_{\rm PS-k}$ both approach the value $\nu
\alpha$. However, in this limit only, the PS-r model is the best 
available model since it directly considers spheres in real space, and
these correspond most directly to the regions that actually collapse
to form halos; this model also applies most directly to the threshold
$\nu$ that is derived based on spherical collapse in real
space. Indeed, the best possible model would be
$\langle\delta(r)\rangle_{\rm ePS-r}$, a quantity based on the
extended Press-Schechter model applied directly in real
space. However, such a model would be extremely difficult to develop,
since excursion sets of $\delta$ as a function of $\sigma^2(r)$ would
no longer correspond to pure random walks [see \citet{bc91} for
related discussion]. Thus, following the tradition of the extended
Press-Schechter approach (see \S \ref{sec:Intro} and
\ref{sec:collapse}), we use $\langle\delta(r)\rangle_{\rm ePS-k}$
along with an ansatz.

To arrive at the most logical ansatz, we simply replace $\alpha$ and
$\beta$ in eq.~(\ref{eq:alpha}) with their $r$-space equivalents.
Based on the above arguments, we adopt $\langle\delta(r)\rangle_{\rm
PS-r}$ and $\Sigma(r)_{\rm PS-r}$ as the correct $\beta
\rightarrow \infty$ limits of the mean and the standard deviation, 
respectively. Indeed, $\delta(r)$ in the ePS-k model approaches a
Gaussian distribution in the limit of $\beta \rightarrow \infty$, and
therefore we can match this limit identically to the PS-r model, and
this procedure fixes unambiguously the required $r$-space versions of
both $\alpha$ and $\beta$. Thus, our final ansats for the mean initial
profile of the density averaged in spheres of radius $r$, linearly
extrapolated to $z=0$, around a virialized halo is
\be \frac{\langle\delta(r)\rangle}{\nu} \equiv \frac{\langle\delta(r)
\rangle_{\rm ePS-k}} {\nu}\ , \label{eq:final1} \ee
where we use eq.~(\ref{eq:ePSk}) except that we define
\be \alpha \equiv  \frac{\xi_r(r_M,r)} {\sigma^2(r_M)}\ \ 
;\ \ \beta \equiv \frac{\nu^2 \alpha (1-\alpha)} {\sigma^2(r)-\alpha\,
\xi_r(r_M,r)}\ . \label{eq:final} \ee We apply these same definitions 
to eqs.~(\ref{eq:Pcum}) and (\ref{eq:Pcum2}) in order to estimate the
scatter in the infall profiles as given by this final model.

\subsection{Halo Virialization in the Presence of Infall}

\label{sec:deltav}

As noted in \S \ref{sec:collapse}, the linearly extrapolated initial
overdensity of a spherical tophat corresponding to the moment of
collapse is \citep{p80} $\del_c = 3 (12 \pi)^{2/3}/20$ if
$\Omega_m=1$. In reality, even a slight violation of the exact
symmetry of the initial perturbation can prevent the tophat from
collapsing to a point. Instead, the halo reaches a state of virial
equilibrium by violent relaxation (phase mixing). Using the virial
theorem $U=-2K$ to relate the potential energy $U$ to the kinetic
energy $K$ in the final state, the final overdensity relative to the
cosmic mean density at the collapse redshift is \citep{p80}
$\Delta_c=18\pi^2$ in the Einstein-de Sitter model. Thus, if a tophat
containing mass $M$ is predicted by spherical collapse to reach a
physical radius $R=0$ at the collapse redshift $z$, this amount of
mass is instead assumed to be distributed within a sphere of size
equal to the (physical) virial radius $R_{\rm vir}$ determined by
$M=\frac{4}{3}\pi \bar \rho(z) \Delta_c R_{\rm vir}^3$.

In the presence of infall, however, the enclosed mass is modified.
Indeed, at the moment when the tophat mass $M$ is predicted by
spherical collapse to reach a radius $R=0$, the surrounding infall
implies that additional mass has already accumulated within the sphere
of radius $R_{\rm vir}$. Thus, the overdensity within the standard
virial radius $R_{\rm vir}$ is predicted to be higher than the
standard value of $18 \pi^2$.  Therefore, once infall is included,
even in the Einstein-de Sitter model the precise overdensity to be
used to define a ``virialized'' halo becomes somewhat uncertain. In
this paper we choose the overdensity of $18 \pi^2$ to {\it define}\/
the virial radius in all cosmological models, since this overdensity
has been frequently used to compare the analytical model predictions
(e.g., for the halo mass function) with numerical simulations
\citep[e.g.,][]{jenk01}. Thus, we carry out spherical collapse calculations
with the average infall profile derived in the previous
section. Having fixed the virial overdensity we then find the enclosed
virial mass $M_{\rm vir}$, which is greater than the original tophat
mass $M$ by up to a factor of two (depending on $M$, $z$, and the
cosmological parameters). We also find the linearly extrapolated
initial overdensity of the virialized mass shell, which we denote
$\del_v$.

In the Einstein-de Sitter case we find $\del_v=1.5927$ for the
infall-based definition of virialization just noted, compared to the
standard value of $\del_c=1.6865$ for an isolated initial tophat
perturbation. More generally, $\del_c$ depends on the cosmological
parameters. In a flat cosmology with $\Omm(z) + \Omega_{\Lambda}(z)=1$
at a redshift $z$, we find the following fitting function accurate to
$5 \times 10^{-5}$ in the range $0.05 < \Omm(z) < 1$ \citep[see also,
e.g.,][]{ecf96}: \be
\del_c=1.6865 \left[1+\frac{x}{1000} \left(4.86-
.51 x -.113 x^2 \right) \right]\ , \label{eq:delc}
\ee
where $x=\log\left[\Omm(z)\right]$. Although $\del_v$ does not
strictly depend only on the cosmological parameters, we find
numerically that it is essentially independent of halo mass. Indeed,
over a wide parameter range of $10^4 M_{\odot} < M < 10^{16}
M_{\odot}$ and $0.05 < \Omm(z) < 1$, $\del_v$ is given to within a
relative accuracy of $3 \times 10^{-4}$ by
\be
\del_v=1.5927 \left[1+\frac{x}{1000} \left(3.9-
.36 x -.09 x^2 \right) \right]\ . \label{eq:delv}
\ee
Thus, for example, $\del_v = 1.573$ for $\Omm(z) = 0.05$, compared to
$\del_c = 1.659$ in this case.

We note that the Press-Schechter halo mass function is known to
underestimate the abundance of rare halos relative to the results of
numerical simulations \citep[e.g.,][]{jenk01}. This can be fixed in
the Einstein-de Sitter model if the collapse threshold is multiplied
by a factor of 0.87, while we find that infall lowers the threshold,
from $\del_c$ to $\del_v$, by a factor of only 0.94. Thus, infall by
itself cannot fully explain the discrepancy in the Press-Schechter
prediction for the number of high-mass halos.

\section{Results}

\label{sec:res}

We illustrate results for the standard cosmological parameters
$\Omega_{m}=0.3$, $\Omega_{\Lambda}=0.7$, $h=0.7$ and a CDM power
spectrum (with initial slope $n=1$) normalized to a variance $\sigma_8
= 0.8$ in spheres of radius 8 $h^{-1}$ Mpc. We consider a wide range
of redshifts but focus on the higher values since at $z < 2$ the
infalling intergalactic gas is very tenuous and may thus be
undetectable; however, note that at low redshift a significant
fraction of the infalling baryons may already be in the form of
virialized dwarf galaxies that are themselves detectable.

\begin{figure}
\includegraphics[width=84mm]{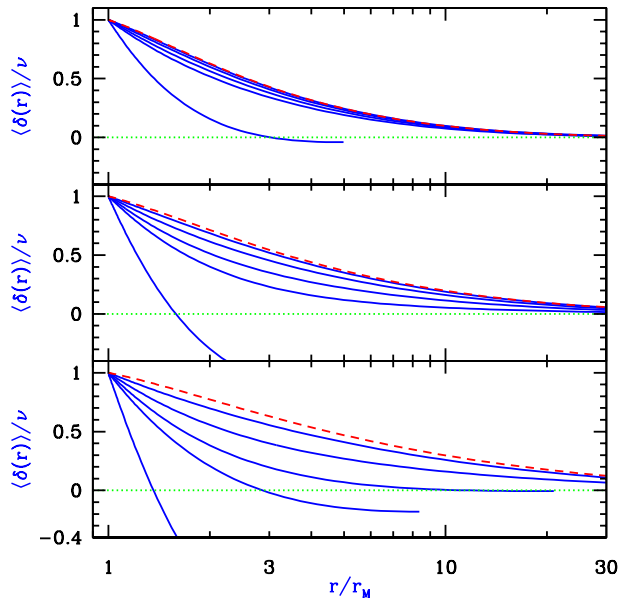} 
\caption{Mean initial density profiles (linearly extrapolated to the
present) around virialized halos. Note that $\delta(r)$ is averaged
over a sphere of comoving radius $r$, and $r_M$ is the initial
comoving radius of the tophat. Results are shown for final halo virial
masses $M_{\rm vir}=10^8 M_{\odot}$ (bottom panel), $10^{10}
M_{\odot}$ (middle panel), and $10^{12} M_{\odot}$ (top panel). In
each case, profiles are shown (solid curves) at $z=0$, 2, 3, 5, and
10, respectively, from bottom to top.  Also shown is the asymptotic
case of an extremely rare halo (dashed curve). The position of zero
overdensity is also indicated (dotted line).}
\label{fig-1}
\end{figure}

Figure~\ref{fig-1} shows the predictions of our model
[eqs.~(\ref{eq:final1}) and (\ref{eq:final})] for the mean initial
profiles of enclosed density around virialized halos. In most cases,
the profiles lie significantly below the asymptotic case of a rare
halo ($\beta \rightarrow \infty$) for which the profile is given by
$\langle\delta(r)\rangle_{\rm PS-r}$ as in eq.~(\ref{eq:PSr}). In
particular, when $M_{\rm vir}=10^8 M_{\odot}$, a halo forming at $z
\la 2$ is most likely to be found in a void on all scales larger than
a few times $r_M$. Note that we restrict all calculations to radii
where the initial profile of enclosed density is monotonically
decreasing, since otherwise shells would cross during the subsequent
collapse; this cutoff affects only the $z=0$ cases as well as the
$M_{\rm vir}=10^8 M_{\odot}$ halo forming at $z=2$ or 3.

\begin{figure}
\includegraphics[width=84mm]{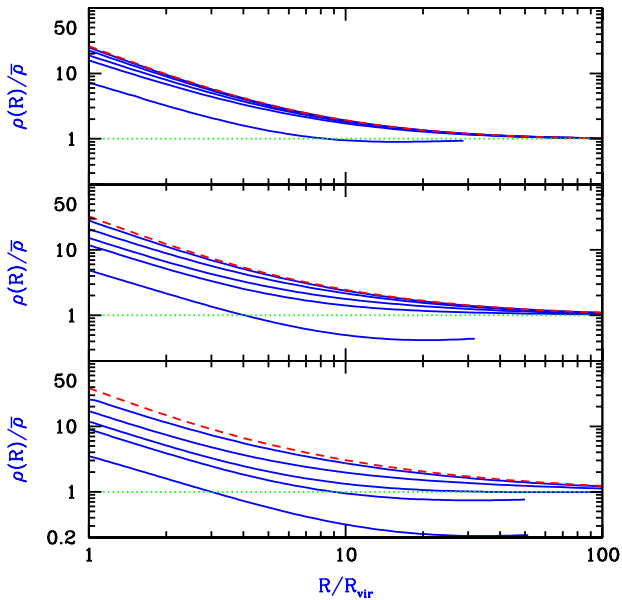} 
\caption{Final density profiles around virialized halos with
the initial profiles shown in Figure~\ref{fig-1}. Note that $\rho(R)$
is the actual density (not enclosed average) at a physical radius $R$,
$\bar{\rho}$ is the cosmic mean density at the virialization redshift,
and $R_{\rm vir}$ is the final virial radius defined as enclosing a
mean density of $18 \pi^2$ times $\bar{\rho}$. Results are shown for
final halo virial masses $M_{\rm vir}=10^8 M_{\odot}$ (bottom panel),
$10^{10} M_{\odot}$ (middle panel), and $10^{12} M_{\odot}$ (top
panel). In each case, profiles are shown (solid curves) at $z=0$, 2,
3, 5, and 10, respectively, from bottom to top.  Also shown is the
asymptotic case of an extremely rare halo (dashed curve). The position
of the cosmic mean density is also indicated (dotted line).}
\label{fig-2}
\end{figure}

We calculate spherical collapse with the initial profiles shown in
Figure~\ref{fig-1}. The resulting final density profiles are shown in
Figure~\ref{fig-2}. In the limiting case of an extremely rare halo
(i.e., very high formation redshift at a fixed halo mass), low mass
halos have higher densities of dark matter surrounding their virial
radius. This can be understood roughly as follows. For a power
spectrum of density fluctuations approximated over some range of
scales as a power law $P(k) \propto k^n$, the correlation function
over the same range of scales varies as $\xi(r) \propto
r^{-(n+3)}$. The initial density profile around a rare halo follows,
roughly, the dark matter correlation function. On small scales, the
CDM power spectrum approaches $n = -3$ and the initial density profile
is rather flat, while on larger scales the power index increases
toward $n = -2$ for large galaxies, which are therefore surrounded by
an initial profile that falls off more steeply (see
Figure~\ref{fig-1}). These differences are magnified by the process of
gravitational collapse, resulting in final values for
$\rho/\bar{\rho}$ at the virial radius of 41, 34, and 27 for $M_{\rm
vir}=10^8$, $10^{10}$, and $10^{12} M_{\odot}$, respectively (see
Figure~\ref{fig-2}). However, at moderate redshifts high mass halos
represent rarer, more extreme peaks than do low mass halos. Thus,
e.g., even at redshift 5 the profile surrounding $M_{\rm vir} \la
10^{10} M_{\odot}$ halos falls significantly below the profile of the
extreme limiting case.

\begin{figure}
\includegraphics[width=84mm]{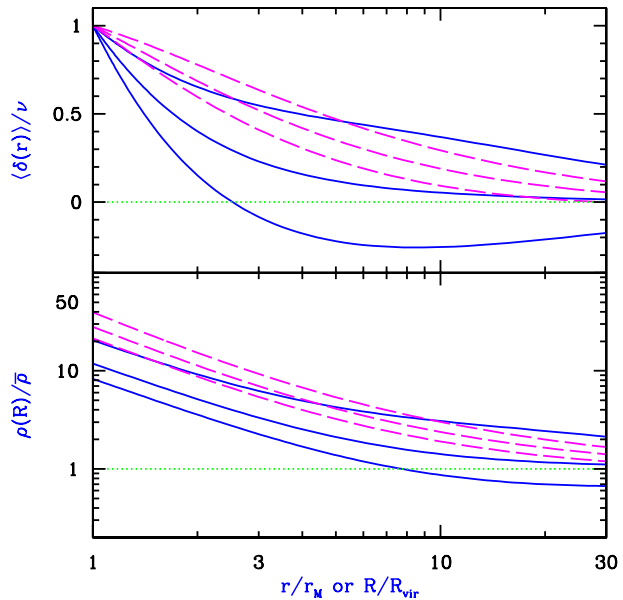} 
\caption{Scatter in the density profiles around virialized halos.
Shown are the initial enclosed density profile (top panel) and the
final density profile (bottom panel), defined as in
Figures~\ref{fig-1} and \ref{fig-2}, respectively. Each curve fixes
the final halo virial mass to $M_{\rm vir}=10^{10} M_{\odot}$, with
the halo forming at $z=2$ (solid curves) or $z=10$ (dashed curves).
In each case, profiles are shown for the minus $1-\sigma$ profile, the
mean expected profile, and the plus $1-\sigma$ profile, respectively,
from bottom to top (see text).}
\label{fig-2p5}
\end{figure}

Next, we use eq.~(\ref{eq:Pcum}) to predict the expected scatter in
the density profiles around virialized halos. Figure~\ref{fig-2p5}
shows the initial and final density profiles for $M_{\rm vir} =
10^{10} M_{\odot}$ at two different redshifts, for the mean expected
profile that we have thus far considered as well as for the plus and
for the minus $1-\sigma$ profiles. The final density at $R_{\rm vir}$,
for example, is higher than the mean density by a factor of
$12^{+8}_{-4}$ at $z=2$ and $28^{+12}_{-7}$ at $z=10$. We thus find a
significant scatter, although it is smaller at the higher redshift. If
the scatter as calculated within our spherical model is representative
of the true scatter among halos, its magnitude suggests that our
predicted mean profile can be compared accurately to an observed
sample if the sample contains at least a few dozen halos of comparable
masses and redshifts.

\begin{figure}
\includegraphics[width=84mm]{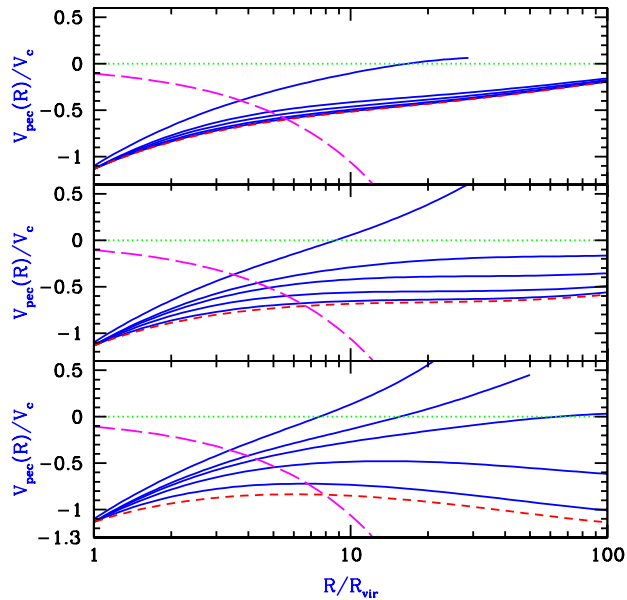} 
\caption{Final velocity profiles around virialized halos with
the initial profiles shown in Figure~\ref{fig-1}. Here $V_{\rm pec}$
is the peculiar velocity (where negative values correspond to infall)
at a physical radius $R$, and $V_c$ is the circular velocity at the
halo virial radius $R_{\rm vir}$. Results are shown for final halo
virial masses $M_{\rm vir}=10^8 M_{\odot}$ (bottom panel), $10^{10}
M_{\odot}$ (middle panel), and $10^{12} M_{\odot}$ (top panel). In
each case, profiles are shown (solid curves) at $z=0$, 2, 3, 5, and
10, respectively, from top to bottom. Also shown is the asymptotic
case of an extremely rare halo (short-dashed curve). For comparison,
the curve corresponding to uniform cosmic expansion is shown (dotted
line), as well as the curve corresponding to stationary
matter (long-dashed curve).}
\label{fig-3}
\end{figure}

Figure~\ref{fig-3} shows the final profiles of peculiar velocity
$V_{\rm pec}$ of the infalling dark matter corresponding to the halos
considered in Figures~\ref{fig-1} and \ref{fig-2}. The same trends are
seen in this figure as in the previous ones; in the extreme case of a
rare halo, infall is stronger for low halo masses, but at moderate
redshifts, the higher-mass halos actually produce the stronger
infall. Figure~\ref{fig-3} also shows the peculiar velocity that would
correspond to stationary matter, i.e., with zero total velocity
(long-dashed curve). This curve is given by minus the Hubble expansion
velocity $V_{\rm Hub}$, which in these units is
\be 
\frac{V_{\rm Hub}(R)}{V_c} = \frac{1}{3 \pi} \frac{R}{R_{\rm vir}}\ .
\ee
Values of $V_{\rm pec}$ that lie below $-V_{\rm Hub}$ give a total
velocity corresponding to infall.

When gas falls into a dark matter halo, incoming streams from all
directions strike each other at supersonic speeds, creating a strong
shock wave. Three-dimensional hydrodynamic simulations show that the
most massive halos at any time in the universe are indeed surrounded
by strong, quasi-spherical accretion shocks \citep{EliSPH,Abel}. The
radius of this shock depends in general on the history of gas infall
into the halo, and will thus vary somewhat among halos depending on
the history of their assembly through mergers. In particular, in the
spherically-symmetric secondary infall solution, when a trace amount
of adiabatic gas is allowed to flow in a gravitational potential
determined by collisionless dark matter with density $\Omm = 1$, the
accretion shock forms at the radius $R_{90.8}$ which encloses a mean
dark matter density of 90.8 times the cosmic mean
\citep{2ndInfall}. However, the simulation of \citet{Abel} shows that
along most lines of sight the shock occurs at a smaller distance,
close to the virial radius. 

The position of the accretion shock is an important theoretical
prediction that can be probed by observations, since the infalling gas
that is about to cross the shock front produces a sharp absorption
feature in high-redshift quasar spectra \citep{nature}. In order to
illustrate the range of possible properties of this gas element that
lies just beyond the shock front, we use the results mentioned above
as a guide and consider two cases for the shock radius $R_{\rm sh}$:
the smaller radius $R_{\rm sh}=R_{\rm vir}$ or the larger radius
$R_{\rm sh}=R_{90.8}$. We assume that the pressure gradient is
negligible in the pre-shock gas compared to gravity, and therefore the
gas density and velocity are equal to those of the dark matter at
distances beyond the shock front.

\begin{figure}
\includegraphics[width=84mm]{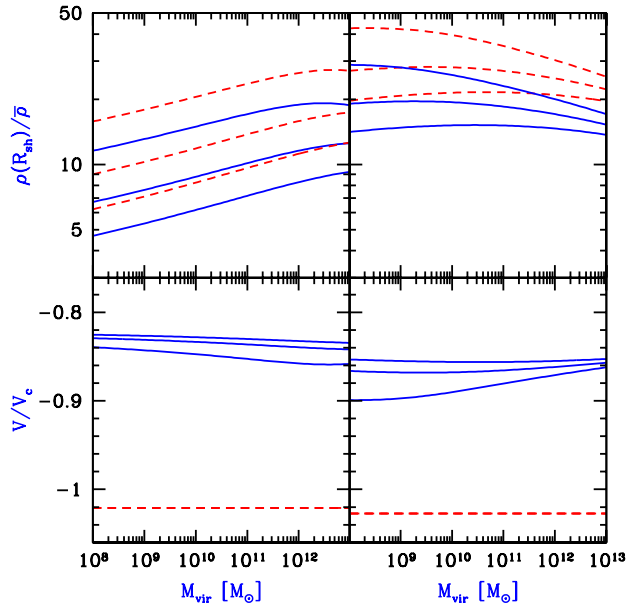} 
\caption{Properties of the infalling gas that is just about to cross
the accretion shock around a virialized halo. Shown are the density in
units of the cosmic mean (top panels), and the total (not peculiar)
velocity in units of the halo circular velocity (bottom panels), both
as a function of the final halo virial mass $M_{\rm vir}$. Results are
shown at $z=2$ (left panels) and $z=10$ (right panels). Two possible
values are considered for the shock radius, $R_{\rm sh}=R_{\rm vir}$
(dashed curves) and $R_{\rm sh}=R_{90.8}$ (solid curves). In each
case, profiles are shown for the minus $1-\sigma$ profile, the mean
expected profile, and the plus $1-\sigma$ profile, respectively, from
bottom to top in the top panels (the order is reversed in the bottom
panels).}
\label{fig-4}
\end{figure}

Figure~\ref{fig-4} shows the density and total infall velocity of the
gas that lies just beyond the shock front. The density and the
magnitude of the infall velocity both tend to increase with redshift,
and follow only weak trends with halo mass at a given redshift. The
differences between the two redshifts shown are fairly significant for
low mass halos but rather small for high mass halos.  In general, the
infalling gas is significantly overdense even for halo masses
corresponding to $z=0$ dwarf galaxies, especially at the high
redshifts which are increasingly becoming the focus of
observations. The figure allows us to compare two sources of scatter,
namely the varying position of $R_{\rm sh}$ and the variability in the
initial density profile. The two effects produce a comparable scatter
in the pre-shock density, while the infall velocity, which depends on
the enclosed mean density rather than on the density itself, is
primarily sensitive to the position of the shock.

\section{Conclusions}

We have developed a model for calculating the initial density profile
around overdensities that later collapse to form virialized
halos. This is the first such model to account for the possibility
that an overdensity on a given scale may be contained inside a large
overdensity on an even bigger scale. As a result, we have found that
the mean expected profile [eqs.~(\ref{eq:ePSk}) and (\ref{eq:final})]
depends on both the mass of the halo and its formation
redshift. Starting from this mean initial profile, we have used
spherical collapse to obtain the final density profile surrounding the
virialized halo.

In reality, there will be some variation in the initial profiles, with
each leading to a different final profile. We have estimated the
scatter within our spherical model, and found that a sample of around
a few dozen halos is required in order to obtain an accurate mean
profile by averaging. Halo samples derived from numerical simulations
or from observations can be used to test our predictions for the mean
and for the scatter of the density and velocity profiles of infalling
matter around virialized halos.

Once infall is included, the standard derivation of the mean density
enclosed within the virial radius fails. Redefining the virial radius
as the radius enclosing a density higher than the cosmic mean by the
standard value of $18 \pi^2$, we have found that the initial
overdensity needed for a halo to form at some final redshift is lower
than it would have to be for an isolated tophat perturbation [compare
eqs.~(\ref{eq:delc}) and (\ref{eq:delv})]. Therefore, accounting for
infall increases the predicted abundance of rare halos for a given
initial power spectrum, but the increase is too small to fully explain
the discrepancy between the Press-Schechter prediction for the number
of high-mass halos and the number seen in numerical simulations.

We have confirmed that rare halos at a given redshift are surrounded
by large, overdense regions of infall, but we have found that the more
numerous halos that correspond to low-sigma peaks are surrounded by
small infall regions that lie within large voids. However, the
infalling gas just beyond the cosmological accretion shock is in
general much denser than the cosmic mean; a density larger by a factor
of at least 10 occurs around the most massive halos at $z \ga
2$. Possible applications of our results include the Lyman-$\alpha$
absorption profiles of high-redshift objects and the proximity effect
of quasars at all redshifts.

\section*{Acknowledgments}
I am grateful for the hospitality of the Institute for Advanced Study,
where this work was begun. I thank Avi Loeb and Evan Scannapieco for
related discussions. I acknowledge the support of an Alon Fellowship
at Tel Aviv University, Israel Science Foundation grant 28/02, and NSF
grant AST-0204514.

\bsp

\label{lastpage}

\end{document}